\documentclass[showpacs,preprintnumbers,amsmath,amssymb,twocolumn,floatfix,nofootinbib]{revtex4}
\usepackage[usenames]{color}
\usepackage{epsfig}
\usepackage{amssymb}

\newcommand{\be}{\begin{equation}}
\newcommand{\ee}{\end{equation}}
\newcommand{\bee}{\begin{eqnarray}}
\newcommand{\eee}{\end{eqnarray}}

\definecolor{grey}{rgb}{0.9,0.9,0.9}
\definecolor{black}{rgb}{0,0,0}

\begin{document}

\title{  Potential dangers when phase shifts are used as a link \\  between experiment and QCD}

\author{A. \v{S}varc}

\address{Rudjer Bo\v{s}kovi\'{c} Institute, \\
Bijeni\v{c}ka c. 54, \\ 
10 000 Zagreb, Croatia\\ 
E-mail: alfred.svarc@irb.hr}
\date{\today}

\begin{abstract}
L\"{u}scher has shown that in single channel problem (elastic region below first inelastic threshold) there exists a direct link between the discrete  value  of the energy in a finite QCD volume and 
the scattering phase shift at  the same energy. However, when the extension of the theorem is made to the baryon resonance sector (multi-channel situation in the inelastic region above first inelastic 
threshold), eigenphases (diagonal multi-channel quantities) replace phase shifts (single channel quantities). It is necessary to stress that the renowned $\pi /2$ resonance criterion is formulated for 
eigenphases and not for phase shifts, so the resonance extracting procedure has to be applied with utmost care. The potential instability of extracting eigenphases from experimental data which occurs 
if insufficient number of channels is used can be reduced if a trace function which explicitly takes multi-channel aspect of the problem into account is used instead of single-channel phase shifts.

\end{abstract} 

\maketitle

As a central task of baryon spectroscopy is to establish a connection between resonant states predicted by  QCD  and hadron scattering  observables, the discovery that QCD can produce a "scattering 
theory" quantity -- phase shift attracted a lot of attention particularly among experimental physicists. L\"{u}scher's theorem \cite{Luscher1, Luscher2} provided this possibility. 
It is well known that resonances do not correspond to isolated energy levels in
the (discrete) spectrum of the QCD Hamiltonian measured  on the  lattice,  so an additional  
effort  is  needed to  extract  resonance parameters  (mass, width, residua/branching fractions) from the "raw" lattice data. In the single-channel case, i.e. in the case of elastic scattering,  the  
pertinent  procedure is   well  known under  the name of L\"{u}scher  framework \cite{Luscher1,Luscher2}. In this framework, for a system  described  by a given  
quantum-mechanical Hamiltonian one relates  the  measured  discrete  value  of the energy in a finite volume to the scattering phase shift at  the  same energy for the  same 
system  in the  infinite volume.  Consequently,  studying  the  volume-dependence of the discrete 
spectrum of the lattice QCD gives the energy dependence of the elastic scattering phase shift and eventually enables one to locate the 
resonance pole positions.

However, as the original  L\"{u}scher's derivation has been done for energies below first inelastic threshold, it was  not directly applicable for scrutinizing baryon spectrum.  
In order to overcome this problem,  this formalism has recently  been generalized to multi-channel scattering and for required baryon resonance energy range.
This was first done in Ref. \cite{Liu} on the basis of potential scattering theory, while later on in Refs. \cite{Lage,Bernard,Doe2012,Doe2011,Doe2011a,Doe2011b,Doe2007,Oset2011}
non-relativistic effective field theory (EFT) have been used for this purpose. Finally, even more general extensions of the theorem beyond a single-channel theory  have also been very recently 
reported  \cite{Ishii2011,HansenSharpe2012}. In all cases conclusions remained very similar as for the single channel case, but with one very important difference. Eigenphases replace phase shifts.  
And it is very important to emphasize that this, seemingly minor change represents a fundamental difference between L\"{u}scher approaches in the elastic, and its generalization to the inelastic 
situation: whereas in the former, one aims at the extraction of a single-channel quantity (the scattering phase shift) which is in principle obtainable from the single-channel measurement, the latter 
case is a multi-channel problem. Not one, but several scattering phases have to be extracted, and scattering matrix diagonalization has to be performed in order to obtain eigenphases. Hence, one has 
to be very careful to apply resonance criteria properly and correctly.

The intention of this letter is  to stress the difference between using phase shifts and eigenphases, and  discuss interrelations among phase shifts, eigenphases, K matrix and T matrix poles as  
potential resonance criteria for quantifying resonance parameters (mass, width, residua/branching fractions). The main purpose is to avoid a confusion and misunderstandings by using physical phase 
shifts instead of eigenphases;  secondary task is to restore the awareness about the importance of a trace function as a tool to remove the instabilities in resonance extraction procedure with 
eigenphases and K-matrix poles by manifestly imposing multi-channel features of a theory.  In spite of looking educational, I believe that this paper is additionally important because it stresses 
principal features of L\"{u}scher approach  and its generalization to inelastic domain with the motive to avoid unjustified simplifications in identifying resonances as has been done in recent, 
renowned experimental work by D\"{u}rr et al. \cite{Durr2008}. In this paper it has been explicitly suggested: 

 \noindent
{ \fontfamily{ppl}\selectfont \small "...The  $\pi \pi $ {\em scattering phase $\delta _{11} (k)$} in the isospin \mbox{I = 1}, spin \mbox{J = 1} channel passes through $\pi /2$ {\em at the resonance 
energy}....",}

 \noindent
so the well-known $\pi /2$ criterion to obtain the resonance mass has been used directly on phase shifts. This is, however, incorrect.
Scattering \emph{eigenphase}, and not \emph{scattering phase} passes through $\pi /2$  at the resonance energy. Single-channel measurement of only one phase shifts is simply not enough, and this 
assumption, even when being fairly reasonable as in the mentioned case, is not generally true.  Instead, one should either use eigenshifts, eigenshift trace or  standard pole determination methods to 
extract T-matrix pole from the energy dependent phase shifts, and not phase shifts directly. Using phase shifts only is erroneous. Therefore, I strongly encourage thorough approaches like it has been 
done in refs. \cite{Lage,Bernard,Doe2012,Doe2011,Doe2011a,Doe2011b,Doe2007,Oset2011} where  L\"{u}scher's formalism has been used to obtain phase shifts, but then an accurate determination of 
resonance pole positions in the multi-channel scattering has been performed..
 
To fulfill the outlined task first brings us to the well known issue of defining what a resonance actually is in scattering theory. A precise definition of a resonance is in principle a nontrivial, 
and even ill defined mathematical problem  \cite{Simon}, but for practical purposes it is sufficient enough to discuss only two alternative definitions as has been suggested by Exner \& Lipovsky in 
\cite{ExnerLipovsky}: we may either define resonances via scattering resonances which are characterized by a prolonged time two particles spend together with respect to the standard scattering  
process\footnote{The lifetime of the particle--target system in the  region of interaction is larger than the collision time in a direct collision process causing a  time delay.}, or through resolvent 
resonances which are characterized by the existence of a pole of the scattering matrix. However, even when these two definitions definitely differ, Exner \& Lipovsky stress that they do coincide for 
most physical situations. So, this allows  us to restrict our discussion to only one of them: we use the existence of scattering matrix poles as a fairly robust criteria for identifying the resonant 
state\footnote{For further reading I recommend  Dalitz-Moorhouse old publication \cite{DalitzMoorhouse}, where  these issues have been extensively elaborated.}.  

In the context of discussing scattering matrix poles, Dalitz \& Moorhouse  have in \cite{DalitzMoorhouse} also  introduced \emph{scattering matrix eigenphases} and extensively discussed the concept 
that the behavior of the resonance eigenphase can be taken as a resonance signal. I quote:

\noindent
{  \fontfamily{ppl}\selectfont \small "... Dalitz (1963)\footnote{See reference \cite{Dalitz63}} and Dalitz \& Moorhouse (1965)\footnote{See  reference \cite{DM65}.} considered the eigenphases $\delta 
_\alpha$ and eigenstates $\phi _\alpha$ of the unitary matrix S, as is certainly always permissible. It then appeared plausible that the (real) \textit{\small resonance energy $E_0$} corresponded to 
one of these \emph{\small eigenphases} increasing rapidly through $\pi /2$."  }

\noindent 
The most important point is that $\pi /2$ resonance criteria for phases is introduced for \emph{eigenphases}, and not for physical channel \emph{phase shifts}. 

In addition to introducing eigenphases as a concept,  they in further analysis also illustrated how this simple  $\pi /2$ criterion actually works in reality, for a multi-channel theory. 
They have shown that multi-channel character and no-crossing theorem  strongly predetermine the delicate behavior of eigenshifts in the vicinity of resonance energy. A simple three channel model with 
constant background phases has been used to show  that $\pi /2$ criterion combined with no-crossing theorem causes that all channels must have a rapid variation near the resonance, but only one of 
them traverses through $\pi /2$. When the energy of the system approaches the resonance value, the first eigenphase experiences a rapid change and approaches the second one. But, instead of crossing 
it and continuing through  $\pi /2$, because of no-crossing theorem it just "bumps" into it and "repels"  transferring the "momentum" to the second phase shift, and continues smoothly on towards the 
constant background value of the second phase shift. The second one, however,  takes over the rapid energy variation and keeps on changing fast. And similar event happens when the second eigenphase 
"meets" the next one. 
 Thus, near the  resonance energy all three eigenphases are required to undergo rapid energy variations over energy ranges small compared with the width r of the resonance at energy $E_0$, but 
actually only one traverses through $\pi /2$. 

This behavior has also been examined in detail by Goebel \& McVoy \cite{Goebel67} and by McVoy \cite{McVoy67}, who show that these rapid energy variations are due to the existence of branch cuts in 
all channel eigenphases $\delta _\alpha (E)$ and corresponding eigenvectors $\phi _\alpha$ on the unphysical sheet of the E plane, and lying much closer to the physical axis than does the resonance 
pole. It is important to notice that these branch cuts do not occur in the complete S matrix $S=\sum _{\alpha} \phi _\alpha e^{2 i \delta_\alpha} \tilde{\phi} _\alpha$ but only in channel eigenphases 
separately, and therefore do not have any physical significance. And the way out has been found by realizing that the only way how this can happen is that the occurrence of these branch cuts in the 
$\phi _\alpha$ and $\delta _\alpha$ must be just such that all these branch cuts exactly cancel out in the full S-matrix combination. Because of that Goebel \& McVoy conclude that with such a 
complexity of branch cuts without physical significance, the eigenphase representation for the S matrix is not generally a useful representation for the scattering in the neighborhood of a resonance.   

Now we are faced with a situation when we have to consider both, poles and eigenphases. 

I believe that four major facts in relating poles and eigenpahses should be stressed: \\
 \textbf{i)} eigenphase $\pi /2$ criterion is equivalent to K matrices having poles at resonant energies (the rapid increase of eigenphase through $\pi /2$  is equivalent to the fact that the 
corresponding eigenvalue of the reaction matrix \mbox{$K = i(S -1)/(S + 1)$} has a pole at this energy, see \cite{Dalitz61}); \\
 \textbf{ii)}  resonance parameters obtained from K and T matrix poles are quantitatively different (in spite of being interrelated  at least for a meromorphic type of background- see Ref. 
\cite{Svarc2012}); \\
 \textbf{iii)} as a direct corrolary of i) and ii) we have to conclude that resonance parameters obtained from eigenphases and from T matrix poles \emph{must be} quantitatively different; 
 and \\
 \textbf{iv)} while the T-matrix poles are in principle single-channel quantities (it is sufficient to measure observables between only one initial and only one final channel to reconstruct the 
T-matrix between these channels), K-matrix  poles and consequently eigenphases are multi-channel quantities (one needs to know reactions between all channels to reliably reconstruct single channel 
K-matrix matrix element as the full coupled-channel T-matrix inverse has to be done)\footnote{An illustration:   one needs to measure  all observables  only for $\pi N \rightarrow \eta N$ reaction in 
order to obtain $\pi N \rightarrow \eta N$ T-matrix, but one needs to measure observables for  all $\pi N \rightarrow X \, Y$  processes to obtain $\pi N \rightarrow \eta N$ K-matrix (inversion of the 
full coupled-channel T-matrix is needed).   Inverting only $\pi N \rightarrow \eta N$ T-matrix gives an incorrect result.}. 

In literature we  usually meet three resonance quantification criteria for resolvent resonances: a pole of the scattering matrix, a pole of the K matrix and the energy when eigenphase increases 
rapidly through  $\pi /2$. However, it is rarely said that second and third criterion \emph{are identical} but \emph{different} from the first one, and  very rarely said that second and third 
criterion tend to be instable if too small number of channels is analyzed. 

Let me now pay some attention to K-matrix and eigenshift instability; to its origin and its implications. 
 
In the case of elastic scattering (single-channel theory) like in the original L\"{u}scher approach,  physical channel phase shift is identical to the S-matrix eigenshift, and single-channel 
measurement suffices. However, for the inelastic region, a multi-channel theory is needed in order to obtain all phase shifts, and physical scattering matrix has to be diagonalized to get eigenphases.  
So, eigenchannels and physical channels differ, and in order to obtain one or all eigenphases one has to know all physical channels at the same time. As a direct consequence of these considerations, 
all criteria formulated on K-matrices and eigenphases tend to be unstable if only one, or too few channels are measured. In other words, while small changes of single channel data can result only in 
small changes of T-matrix poles (T-matrix poles being single-channel quantity), small changes of single-channel data can indeed produce big changes of K-matrix poles and eigenphases, since other 
non-observed hence not controlled channels can be drastically different. So,  in matrix inversion procedure for obtaining K matrix, or in diagonalization procedure to obtain eigenshifts, notable 
changes in individual members can be introduced even when one channel is kept almost fixed.

This instability, and the multi-channel feature of eigenphases was the main reason why a trace function (in this particular case eigenphase trace) have been introduced. Namely, as it has previously 
been stated,  Goebel \& McVoy \cite{Goebel67} and McVoy \cite{McVoy67} have demonstrated that the individual branch cuts in each channel eigenphase must exactly cancel out in the full S-matrix, so a 
trace of eigenphases being a sum of eigenphases must also be free of these individual branch cuts. Following old Macek 1970 idea \cite{Macek1970},  U. Hazi has explicitly shown \cite{Hazi1979} that 
for an isolated resonance in a multichannel problem the sum of the eigenphases $\delta _\alpha$ (eigenphase trace), and not individual eigenphases satisfies the usual formula appropriate for the 
elastic phase shift: $\rm{tr} \, (\delta _\alpha) = \Delta _0 + tan^{-1} \, [r/2(E_0 - E)]$ where $\Delta _0$ is the sum of background phases. This sum (the trace) explicitly enforces multi-channel 
character of the problem, so standard techniques used for phase shifts in a single-channel theory can be explicitly used for eigentrace in a multi-channel theory. This feature has also been explicitly 
discussed in recent Ceci. et al reference \cite{Ceci2009} where it has been demonstrated that a K-matrix trace can be used to relate K-matrix poles and standard T-channel Breit-Wigner parameters in a 
background independent way.

These issues have been recently recognized by several groups, and each of them offered its own way to overcame the problem. 

One of them is the GWU group \cite{GWU2012} where the authors have analyzed the use influence of different K-matrix parametrization on eigenphases and T-matrix poles. The authors have shown that 
regardless whether Chew-Mandelstam K-matrix is parameterized either in a form of polynomial, or in a form of poles with nonsingular background, T-matrices are very similar. However, they show that 
eigenphases, are very different. It is very important that  they are able to relate the origin of this difference to the fact that they fit only $\pi N$ elastic and $\eta$ production channel, so 
uncertainties in other channels cause eigenphases (and K-matrix poles consequently) to vary. They also introduce the trace function (but not for eigenphases but for their derivatives), and demonstrate 
its advantages over individual channel quantities.

The second group is the Bonn-J\"{u}lich-Valencia collaboration, where they have used a framework based on unitarized
Chiral Perturbation Theory (UCHPT) for the extraction of the scalar resonance parameters.  This model was very successful in the infinite volume, and reproduced well the $\pi \pi /\pi \eta$ and $K 
\bar{K}$ data up to 1200 MeV \cite{Bernard}. Later on it was also  extended to the finite volume considerations \cite{Doe2011b} with considerable success.  The most important point of all is that they 
recognize the fact that $\pi /2$ resonance criteria can not be used to extract pole positions, but they extract them directly from the T-matrix poles. They address two main issues. The first one is 
the use of fully relativistic propagators in the effective field theory framework in a finite volume, and the second one is to discuss in detail the analysis of "raw" lattice data for the 
multi-channel scattering.  They supplement lattice data by a piece of the well-established prior phenomenological knowledge that stems from UCHPT, in order to facilitate the extraction of the 
resonance parameters. In particular, they show that, with such
prior input, e.g., the extraction of the pole position from the data corresponding only to the periodic boundary conditions, is indeed possible. In order to verify the above statements, they  analyze 
"synthetic" lattice data. To this end, they produce energy levels by using UCHPT in a finite volume, assume Gaussian errors for each data point, and then consider these as the lattice data, forgetting 
how
they were produced (e.g., forgetting the parameters of the effective chiral potential and the value of the cutoff). In the analysis of such synthetic data, they test their approach, trying to 
establish resonance masses and widths as scattering matrix poles from the fit to the data.

As only two 2-body channels are nowadays fairly well known ($\pi N$ elastic scattering and $\eta$ production), the use of trace formalism is unfortunately practically impossible. Consequently, using 
trace function is rather neglected, and single channel K-matrices or single-channel eigenphases are very often erroneously used instead of  K-matrix and eigenphase traces. This, however, only stresses 
the critical lack of experimental data in inelastic channels, and shows that  new measurements of all possible hadronic reactions in baryon resonance energy range 1.5 GeV $\leq$ E $\leq$ 2.5 GeV are 
badly needed.  So I strongly endorse a new proposal for J-PARC experiment at 50 GeV Proton Synchrotron \cite{Hicks2012}.

As a summary I would just like to remind the physics community that using L\"{u}scher's theorem to establish a connection between QCD and experiment via phase shifts has to be done with care in real 
baryon resonance energy range. Eigenphases (diagonal multichannel and not single channel quantities) replace phase shifts, so the well-known $\pi /2$ criterion to obtain the resonance mass can not be 
used directly on phase shifts as it has been suggested in a well known D\"urr et al paper \cite{Durr2008}. I would also like to stress the importance  of using traces instead of using single channel 
quantities in case when K-matrices or eigenphases are analyzed, as delicate cancellations are needed  to remove the influence of individual branch cuts in each channel separately 
\cite{DalitzMoorhouse}.  Single channel analysis for K-matrix matrix elements or eigenphases should be by all means avoided, a trace function (basically a multi-channel quantity) should be used 
instead.

\end{document}